
%
%
%
%
%
\magnification=\magstep1
\voffset=-0.65 true cm \hoffset=-1.4 true cm
\vsize=27.0 true cm \hsize=18.5 true cm
\nopagenumbers
\font\mittel=cmbx10 scaled\magstep2
\def\shiftleft#1{#1\llap{#1\hskip 0.04em}}
\def\shiftdown#1{#1\llap{\lower.04ex\hbox{#1}}}
\def\thick#1{\shiftdown{\shiftleft{#1}}}

\def\b#1{\thick{\hbox{$#1$}}}
%
%
\vbox{\vskip 3.1 true cm}
\leftline{\hskip 2.8 truecm
{\bf \mittel Meson-meson correlations in baryon-baryon
} }
\smallskip
\leftline{\hskip 2.8 truecm
{\bf \mittel and antibaryon-baryon interactions
} }
\vskip 0.75 truecm
\leftline{\hskip 2.8 truecm K. Holinde}
\par \bigskip
\leftline{\hskip 2.8 truecm {\sl Institut f\"ur Kernphysik (Theorie),
Forschungszentrum J\"ulich GmbH,} } \par \smallskip
\leftline{\hskip 2.8 truecm
 {\sl  D-52425 J\"ulich, Germany} }
\vskip 2.5 true cm
\leftline{\hskip 2.8 truecm  {\bf ABSTRACT} } \medskip
\noindent
Recent work of the J\"ulich group about the role of meson-meson
correlations in baryon-baryon and antibaryon-baryon interactions
is reviewed.
\bigskip \leftline{\hskip 2.8 truecm  {\bf KEYWORDS} } \medskip
\noindent
Meson-meson correlations, baryon-baryon interactions, antibaryon-baryon
interactions.
\bigskip \medskip
\leftline{\hskip 2.8 truecm  {\bf INTRODUCTION} } \medskip
\noindent
Quantum chromodynamics (QCD) is the underlying theory of strong
interactions, with quarks and gluons as fundamental degrees of
freedom. However, in the non-perturbative region of low and medium
energy physics, mesons and baryons definitely keep their importance
as efficient, collective degrees of freedom for a wide range of
nuclear phenomena. One possibility to detect quark effects is to
treat as many hadronic reactions as possible from a combined
conventional viewpoint, in terms of baryons and mesons. Only by
such an elaborate and consistent treatment one will be able to
reliably explore the limits of the conventional picture and hopefully
establish discrepancies with the empirical situation, which, in turn,
can possibly be identified with explicit quark-gluon effects.

\noindent
In this talk I want to demonstrate the outstanding role of meson-meson
correlations in baryon-baryon interactions. First we will determine
correlated $\pi \pi$ as well as $K \bar K$- exchange between two
baryons, which
should replace the sharp mass $\sigma'$ and $\rho$ exchange used before
in the Bonn nucleon-nucleon ($NN$) [1] and J\"ulich hyperon-nucleon [2]
interaction models. We will then describe the effect of correlated
$\pi \rho$ exchange in the $NN$ interaction, especially its implication
for the (strong) $\pi NN$ formfactor and the nature of the short range
$NN$ repulsion. Finally we will look at the role of both correlations
in nucleon-antinucleon ($N \bar N$) annihilation.
\bigskip
\medskip
\goodbreak
\noindent
\leftline{\hskip 2.8 truecm  {\bf THE BONN NN POTENTIAL } }\medskip
\noindent
Starting point are baryon-baryon-meson vertex functions (Fig.1)
involving not
only nucleons but also $\Delta$-isobars, and various mesons below
1 GeV. The analytic structure of these couplings is (essentially)
determined by the quantum numbers of particles involved at the vertex.
The strength is parametrized by coupling constants; in addition form
factors
are included, which in principle depend on all four-momenta of the
particles involved at the vertex, and are normalized to 1 when all
parameters are on the mass shell.

\vfill
\phantom{ }
\vskip 2.4cm
\includegraphics{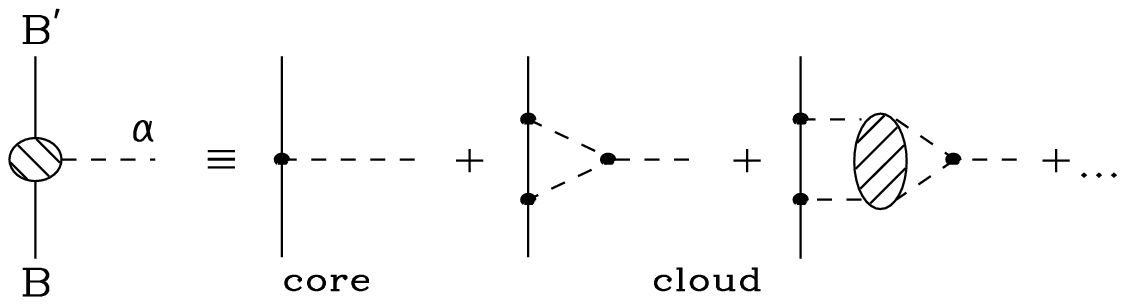}
\hskip2.8truecm{\noindent \sl Fig.1: Baryon-baryon-meson vertex functions}
\vskip0.5cm
\noindent
These vertex functions should be considered as 'elementary' building
blocks for an effective theory of nuclear reactions. As such they
represent a link between QCD and nuclear physics. Ultimately therefore,
QCD should determine the values of the coupling constants and the
structure of the various form factors. Corresponding calculations are
just beginning to emerge. For the $\pi NN$ vertex, with nucleons
restricted
to their mass shells, a recent QCD lattice calculation [3] yields
indeed a reasonable value for $g_{NN \pi} = 12.7 \pm 2.4$ and a
monopole
structure of the form factor with the cutoff mass
$\Lambda_{NN \pi} = 0.75 \pm 0.14$ GeV.

\noindent
In the Bonn potential [1] all form factors are parametrized in a
monopole or dipole form with adjustable cutoff masses $\Lambda_{B'B\alpha}$,
\hbox{$F_{B'B\alpha}=\left( {\Lambda^2-m_{\alpha}^2}\over
{\Lambda^2+\vec q_{\alpha}^2} \right)^n$, n=1,2}.
This procedure provides a total of about
10 free
parameters (coupling constants and cutoff masses), which have been
fixed
by a fit to all $NN$ data below pion production. These vertex functions
should then be used consistently in other hadronic reactions in order
to
explore the limits of the meson exchange concept.

\noindent
Based on these vertex functions a nucleon-nucleon potential has been
constructed (Fig.2) which contains apart from single meson exchanges
higher
order processes involving crossed two-meson exchanges and the
$\Delta$-isobar in intermediate states. Correlated two-pion exchange
(Fig.2d)
is approximated by sharp mass $\sigma'$ and $\rho$ exchange.

\phantom{ }
\vskip 3cm
\includegraphics{fig2.epsi}
\hskip2.8truecm{\noindent \sl Fig.2: Processes included in the Bonn potential}
\vskip0.5cm
\noindent
The resulting description of the $NN$ data is very good; the achieved
$\chi^2$ value per observable of less than 2 [4] certainly proves that
the meson exchange picture essentially works in the low energy $NN$
system.
Furthermore the Bonn potential appears to be a realistic starting point
for finite nuclear calculations. Of course questions remain, e.g.
about the true nature of $\omega$ exchange. It is well conceivable that
$\omega$ exchange as used in the Bonn potential consists only partly of
true $\omega$~exchange but parametrizes in addition different dynamics
arising from higher order meson exchange diagrams and possibly explicit
gluon exchange processes. We will come back to this point later.
\bigskip
\medskip
\goodbreak
\noindent
\leftline{\hskip 2.8 truecm  {\bf CORRELATED \b{\pi\pi + K \bar K}
EXCHANGE  } }\medskip
\noindent
In this chapter I want to describe briefly our dynamical model [5,6]
for correlated two-pion and two-kaon exchange in the baryon-baryon
interaction (Fig.3), both in the scalar-isoscalar ($\sigma$) and
vector-isovector ($\rho$) channel. The contribution of correlated $\pi
\pi$ and $K \bar K$ exchange is derived from the amplitudes for the
transition of a baryon-antibaryon state ($B \bar B'$) to a $\pi \pi$
or $K \bar K$ state in the pseudophysical region by applying
dispersion theory and unitarity. For the $B \bar B' \rightarrow \pi
\pi, K \bar K$ amplitudes a microscopic model is constructed, which is
based on the hadron exchange picture.
\vfill \eject
\vfil
\phantom{ }
\vskip 2.9cm
\includegraphics{fig3.eps}
\hbox to \hsize{\hskip2.8truecm{\noindent
\sl Fig.3: }\vtop{\advance\hsize by -5.6truecm\advance\hsize by -30pt
\noindent \sl Correlated $\pi \pi$ and $K \bar K$ exchange}\hfill}
\vskip0.5cm
\noindent
The Born terms include contributions
from baryon exchange as well as $\rho$-pole diagrams (Fig.4). The
correlations between the two pseudoscalar mesons are taken into
account by means of a coupled-channel ($\pi \pi, K \bar K$) model
[10,11] generated
from s- and t-channel meson exchange Born terms (Fig.5). This model
describes the empirical $\pi \pi$ phase shifts over a large energy
range,
from threshold up to 1.3 GeV. [Quite naturally in this model, the
$f_0$(980)
turns out to be a $K \bar K$ bound state. The genuine scalar resonance
$\varepsilon$
has its physical mass at a much higher energy, about 1.3 GeV, and might
be identified with the chiral partner of the pion.] The parameters of
the $B \bar B' \rightarrow \pi \pi, K \bar K$ model, which are related
to each other by the
assumption of SU(3) symmetry, are determined by the adjustment to the
quasiempirical $N \bar N \rightarrow \pi \pi$ amplitudes in the
pseudophysical region,
$t \ge 4 m_{\pi}^2$ [7] obtained by analytic continuation of empirical
$\pi N$
and $\pi \pi$ data.

\phantom{ }
\vskip 2.4cm
\includegraphics{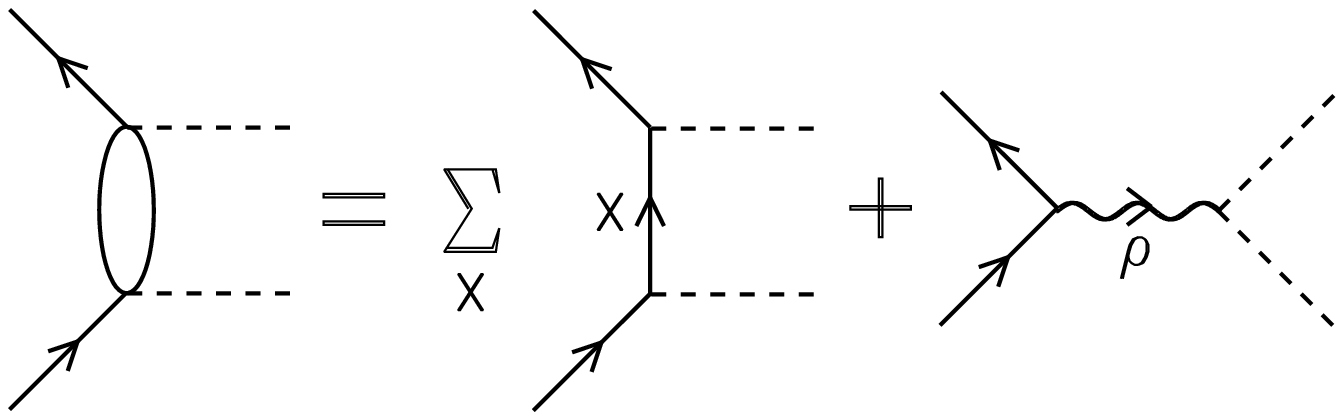}
\hbox to \hsize{\hskip2.8truecm\noindent
{\sl Fig.4: }\vtop{\advance\hsize by -5.6truecm\advance\hsize by -30pt
\noindent \sl Microscopic model for the
$B \bar{B'}\rightarrow \pi \pi, K \bar{K}$
Born amplitudes.
The solid lines denote (anti-)baryons, the dashed lines the
pseudoscalar mesons  $\pi\pi$ or $K \bar{K}$.
The sum over exchanged baryons $X$ contains all members of the
$J^P={1\over2}^+$ octet and the $J^P={3\over2}^+$ decuplet which can
be exchanged in accordance with the conservation of strangeness and
isospin.}\hfill}
\vskip0.5cm

\noindent
Schematically the baryon-baryon potential due to correlated $\pi \pi$
and $K \bar K$ exchanges can be represented as
$$
V^{(0^+,1^-)}_{B'_1,B'_2;B_1,B_2} (t) \sim
\int\limits^{\infty}_{4 m_{\pi}^2} dt' \,
{ {\rho^{_(0^+,1^-)}_{B'_1,B'_2;B_1,B_2} (t')} \over {t'-t} }, \; t<0
\eqno(1)
$$
in terms of spectral functions, which characterize both the strength
and the range of the interaction. Clearly, for sharp mass exchanges,
the spectral function becomes a delta-function at the appropriate
mass.

\phantom{ }
\vskip 2.4cm
\includegraphics{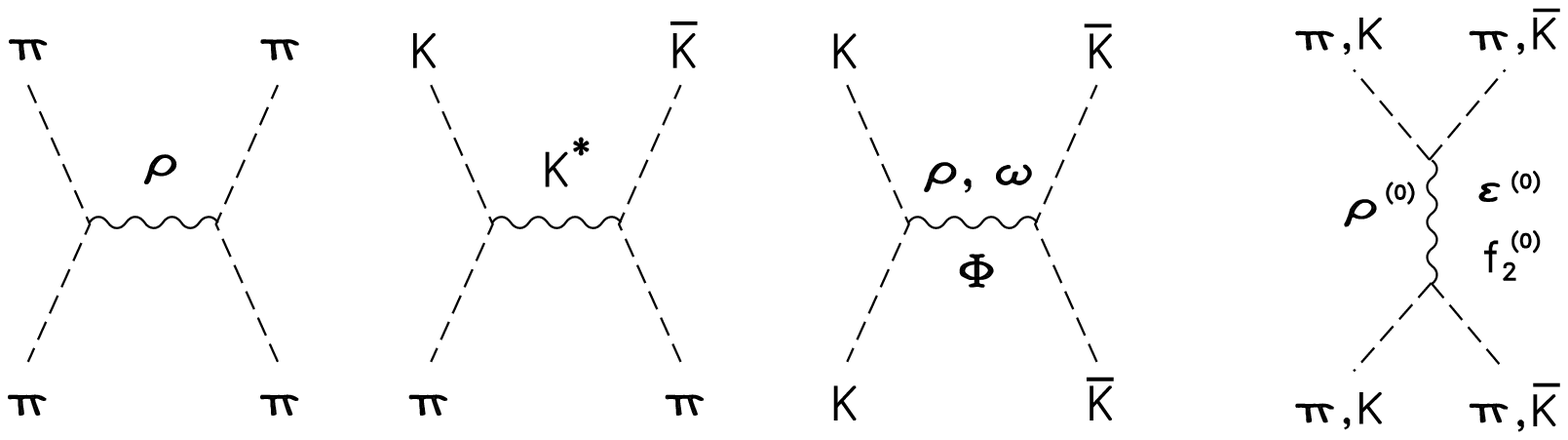}
\hbox to \hsize{\hskip2.8truecm\noindent
{\sl Fig.5: }\vtop{\advance\hsize by -5.6truecm\advance\hsize by -30pt
\noindent \sl Born amplitudes included in the model of Refs.[10,11]
for the $\pi\pi-K \bar{K}$ interaction.}\hfill}
\vskip0.5cm

\noindent
Fig.6 shows the resulting spectral functions in the scalar-isoscalar
channel for the $N N \rightarrow N N$ and
$\Sigma N \rightarrow \Sigma N$ reaction compared to
sharp mass $\sigma$ exchange used in the Bonn $NN$ and J\"ulich $YN$
($Y = \Lambda, \Sigma)$ models.
Obviously the $\Sigma N$ spectral function is
considerably smaller than the result for the $NN$ case. Note also that
correlated $K \bar K$ exchange processes play a minor role in the
$NN$ system
but are very important for the interactions involving hyperons.
Compared to sharp-mass $\sigma'$ exchange of the Bonn potential the
present dynamical treatment provides stronger attraction. The
difference
is especially large in high partial waves since $\sigma'$ exchange does
not contain the long-ranged part of the correlated processes. As
demonstrated in Ref. [5] our result indeed requires a lower
$\pi NN$ coupling constant in order to accommodate the empirical high
partial wave $NN$ scattering phase shifts, as suggested for quite some
time by the Nijmegen group [8].

\phantom{ }
\vskip 5cm
\includegraphics{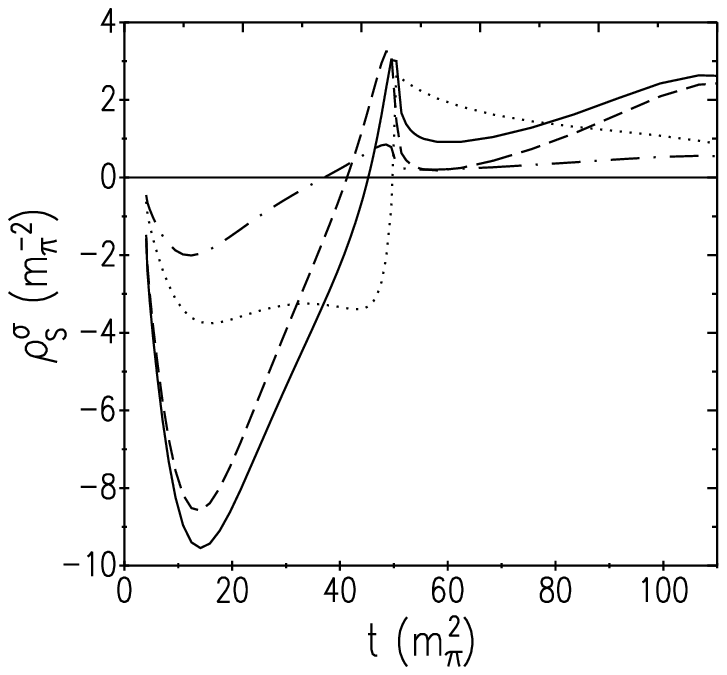}
\hbox to \hsize{\hskip2.8truecm {\noindent
\sl Fig.6: }\vtop{\advance\hsize by -5.6truecm\advance\hsize by -30pt
\noindent \sl Spectral function $\rho^\sigma_S(t)$
in the $NN$ (solid) and $N \Sigma$ (dotted) channel derived with the
full model (cf.\ Fig.3).
If the contributions of the
 $B \bar {B'}\to K \bar {K}$ Born amplitudes are neglected the dashed
($NN$) and the dash-dotted lines ($N\Sigma$) are obtained.}\hfill}
\vskip0.5cm

\noindent
For the following discussion it is convenient to parametrize our
results in terms of effective coupling strengths. In the simple
one-boson-exchange picture the intermediate range attraction is
provided by $\sigma$ exchange; the underlying baryon-baryon potential
has the following structure
$$
V^{(\sigma)}_{B'_1,B'_2;B_1,B_2} (t) \sim
g^{\sigma}_{B'_1 B_1} g^{\sigma}_{B'_2 B_2} {
{1} \over {m_{\sigma}^2-t}} \; .
\eqno(2)
$$
The correlated potential is given by Eq. (1), which can be
parametrized
in terms of t-dependent strength functions
$G^{(0^+)}_{B'_1,B'_2;B_1,B_2}(t)$.
$$
V^{(0^+)}_{B'_1,B'_2;B_1,B_2} (t) \sim
G^{(0^+)}_{B'_1,B'_2;B_1,B_2}(t)  {{1} \over {m_{\sigma}^2-t}}
\eqno(3)
$$

\noindent
Note that in the OBE frame the three reactions $N N \rightarrow NN$,
$\Sigma N \rightarrow \Sigma N$,
$\Sigma \Sigma \rightarrow \Sigma \Sigma$
are determined by two parameters (coupling constants) $g^{\sigma}_{NN}$
and $g^{\sigma}_{\Sigma \Sigma}$
whereas the correlated exchanges are characterized by three independent
strength functions, which means of course that vertex coupling
constants
cannot be well defined.

\noindent
In the physical region the strength of the contributions is to a large
extent governed by the value of G at t=0. The values for the various
channels (with normalization 1 in the $NN$ channel) are shown in Table 1.
%
%
\midinsert
\hbox to \hsize{\hskip2.8truecm\noindent
{\sl Table 1: }\vtop{\advance\hsize by -5.6truecm\advance\hsize by -30pt
\noindent \sl Values of the strength function $G^{(0^+)}(t=0)$ for $BB'$
channels
with increasing strangeness. }\hfill}
$$
\vbox{
{\offinterlineskip \tabskip=0pt
\halign{  \strut
          #\quad&
          #  &
          #&
          \quad\hfill #\quad &
          #&
          \quad\hfill #\quad &
          #&
          \quad \hfill # \quad &
          #&
          \quad \hfill # \quad &
          #&
          \quad \hfill # \quad &
          #&
          \quad \hfill # \quad &
          #\cr
\noalign{\hrule}
&    && $NN$ && $\Lambda N$ && $\Sigma N$ && $\Lambda \Lambda$ &&
        $\Sigma \Sigma$ && $N \Xi$ & \cr
\noalign{\hrule}
& $G^{(0^+)}(t=0)$ && 1\phantom{.0} && 0.49 && 0.41 && 0.26 && 0.30 &&
                      0.19 & \cr
\noalign{\hrule}
}}
}
$$
\endinsert
%
%
\noindent
Thus as a main result the strengths of correlated $\pi \pi$ and
$K \bar K$
exchange decrease with increasing strangeness (becoming more
negative). Furthermore they do not fulfill SU(3) relations.

\noindent
The situation is similar but more complicated in the vector-isovector
($\rho$) channel. Now we have in the OBE at each vertex a vector (g)
and
a tensor (f) coupling. Thus the $NN$ interaction is determined by
two coupling constants, $g_{\rho NN}$ and $f_{\rho NN}$ whereas the
correlated result requires
three strength functions $G^{vv}_{NN,NN}(t)$, $G^{tt}_{NN,NN}(t)$,
$G^{vt}_{NN,NN}(t)$. Fig.7 shows our
results in comparison to the corresponding coupling constants
including formfactors used in the Bonn potential. Again our results
are considerably larger. The reason is the form factor with
$\Lambda_{NN \rho}$ = 1.4 GeV used in the Bonn potential, which
reduces the strength at t=0 by as much as 50 \%.

\phantom{ }
\vskip 7.5cm
\includegraphics{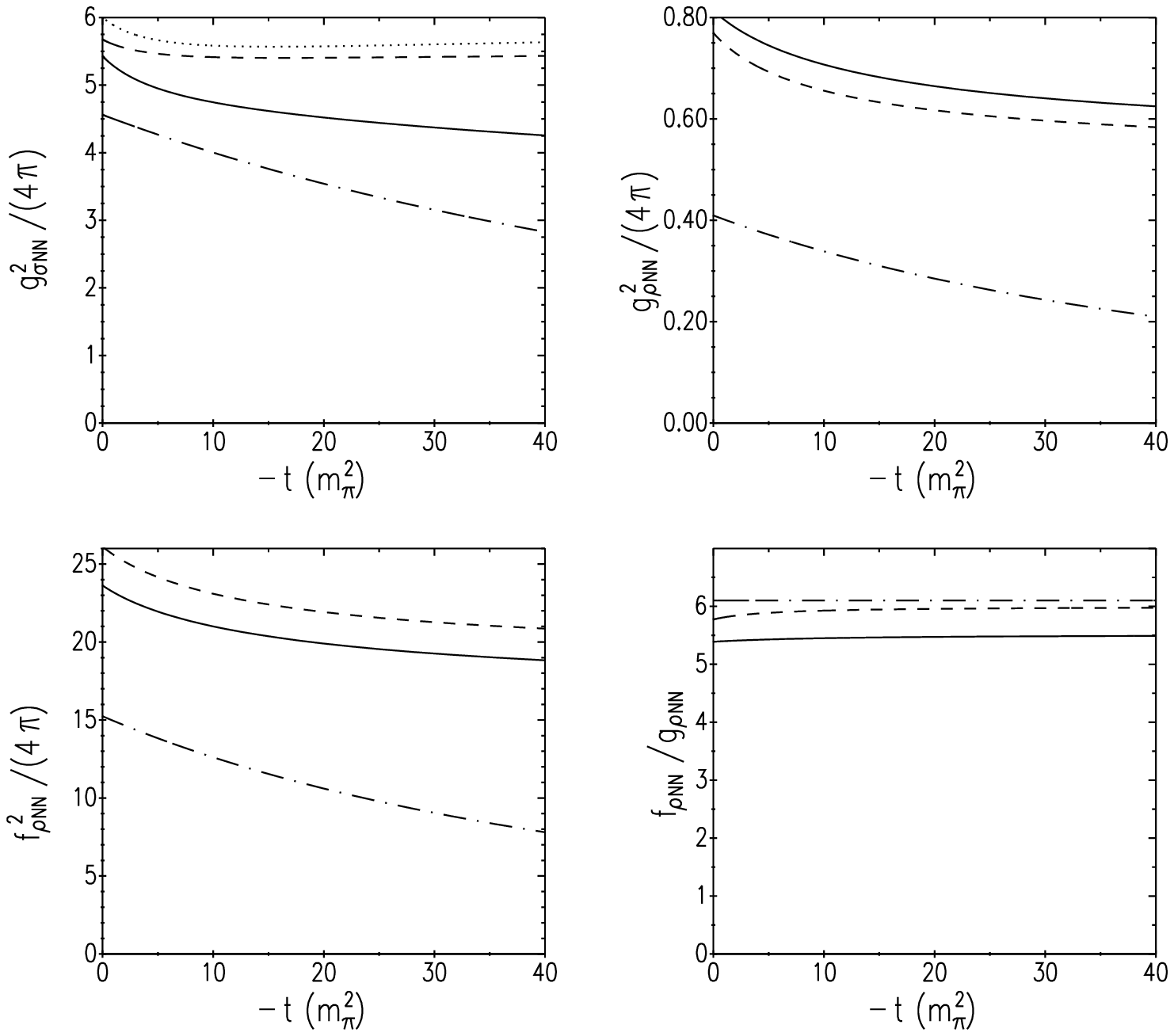}
\hbox to \hsize{\hskip2.8truecm {\noindent
\sl Fig.7: }\vtop{\advance\hsize by -5.6truecm\advance\hsize by -30pt
\noindent \sl Effective  strength of the  $NN$ interaction due to
correlated $\pi\pi$ and $K \bar {K}$ exchange  in the  $\sigma$ and
$\rho$ channel as a function of the 4-momentum transfer $t<0$.
Shown are
$g^2_{\sigma NN}\!\equiv\! G^\sigma_{\scriptscriptstyle NN \to NN}$,
$g^2_{\rho NN}\!\equiv\, {^{\scriptscriptstyle VV}G^\rho_{\scriptscriptstyle NN
\to NN}}$,
$f^2_{\rho NN}\!\equiv \, {^{\scriptscriptstyle TT}G^\rho_{\scriptscriptstyle
NN \to NN}}$ and
$f_{\rho NN}/g_{\rho NN} \!\equiv \!
[{{^{\scriptscriptstyle TT\!}G^\rho_{\scriptscriptstyle NN \to NN}}/\,
{^{\scriptscriptstyle VV\!}G^\rho_{\scriptscriptstyle NN \to NN}}}]^{1\over
2}$.
The solid (dotted) line is derived from  the microscopic model for
correlated
$\pi\pi$ and $K \bar {K}$ exchange using $t'_{max}=120m_\pi^2$
($t'_{max}=50m_\pi^2$) as cutoff in the dispersion integral.
The dashed line follows from the  quasiempirical
$N \bar N\to\pi\pi$
amplitudes [7] with $t'_{max}=50m_\pi^2$.
The effective strength of $\sigma'$ and $\rho$ exchange in the Bonn
potential[1]
is denoted by the dash-dotted line.}\hfill}
\vskip0.5cm

\noindent
 For the general baryon-baryon case, the resulting strength
functions can be likewise determined. Again they deviate from what
is expected in the naive SU(3) picture for genuine $\rho$ exchange, due
to a sizable admixture of baryon exchange processes to the (SU(3)
symmetric) $\rho$-pole contributions.
\bigskip
\medskip
\goodbreak
\noindent
\leftline{\hskip 2.8 truecm  {\bf CORRELATED \b{\pi\rho} EXCHANGE
 } }\medskip
\noindent
The importance of $\rho$-exchange for the dynamics of the $NN$ system
derives from the following fact: It provides a sizable intermediate
range tensor force, which has opposite sign to the tensor force
generated by one-pion exchange. Thus there is a strong
cancellation, over a relatively broad range of energies and distances,
between $\pi$ and $\rho$ exchange in the tensor channel. A similar
cancellation occurs between K and K* exchange, e.g. in the
hyperon-nucleon interaction. Therefore, in the $NN$ (and in the
baryon-baryon
system in general) it is strongly suggested to always group $\pi$ and
$\rho$
(as well as K and K*) together in order to reach sufficient convergence
in the expansion of the irreducible kernel (potential).

\noindent
In fact this procedure has been an essential guideline when
constructing
the Bonn potential. Unfortunately it was not followed to a sufficient
degree: Whereas, in second order diagrams (cp.\ Fig.2) $\pi \pi$ as
well
as
$\pi \rho$ exchange has been included for uncorrelated processes (with
N
and $\Delta$ intermediate states) this has not been done for correlated
processes: correlated $2 \pi$-exchange processes have been effectively
included (in terms of sharp mass $\sigma'$ and $\rho$ exchange) but
correlated $\pi \rho$ processes have been left out. The reason is quite
simple: The evaluation of this missing piece is technically quite
complicated, much more involved (due to the spin of the $\rho$)
compared
to correlated $2 \pi$-exchange. More importantly, a dynamical model for
the
interaction between a $\pi$ and a $\rho$ meson was not available at
that time.

\phantom{ }
\vskip 6cm
\includegraphics{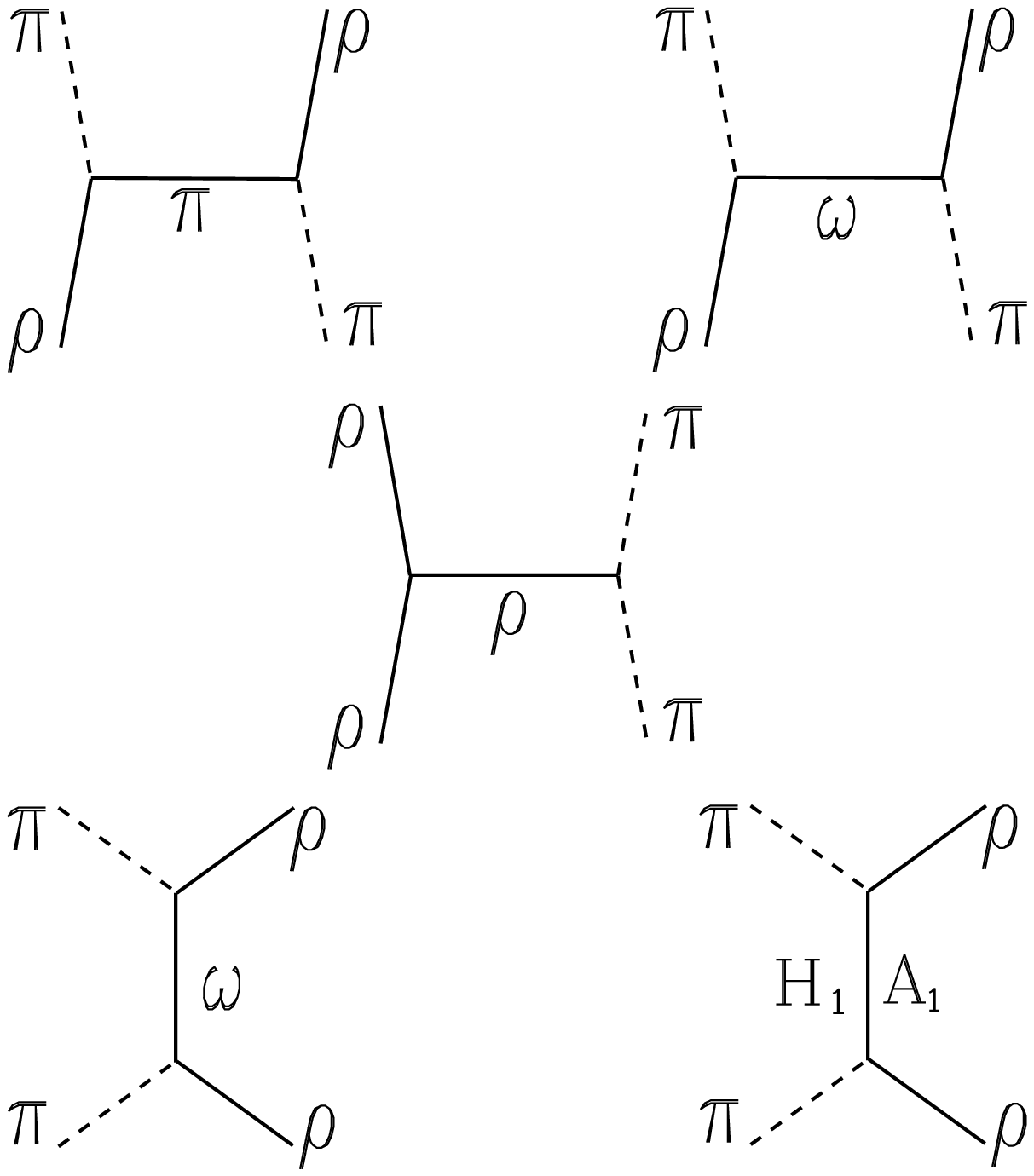}
\hbox to \hsize{\hskip2.8truecm{\noindent
\sl Fig.8: }\vtop{\advance\hsize by -5.6truecm\advance\hsize by -30pt
\noindent \sl Born terms included in the model of Ref.[9] for the
$\pi\rho$ interaction.}\hfill}
\vskip0.5cm

\noindent
We have recently constructed a corresponding potential model [9],
with
driving terms shown in Fig.8. Open parameters have been adjusted
to empirical pole parameters of the $A_1$, $H_1$, and $\omega$ meson.

\noindent
This $\pi \rho$ T-matrix is now inserted into the correlated $\pi \rho$
exchange
diagram of Fig.9. The evaluation proceeds via the same
dispersion-theoretical treatment as used for the $\pi \pi$ case in the
foregoing section.

\phantom{ }
\vskip 3.0cm
\includegraphics{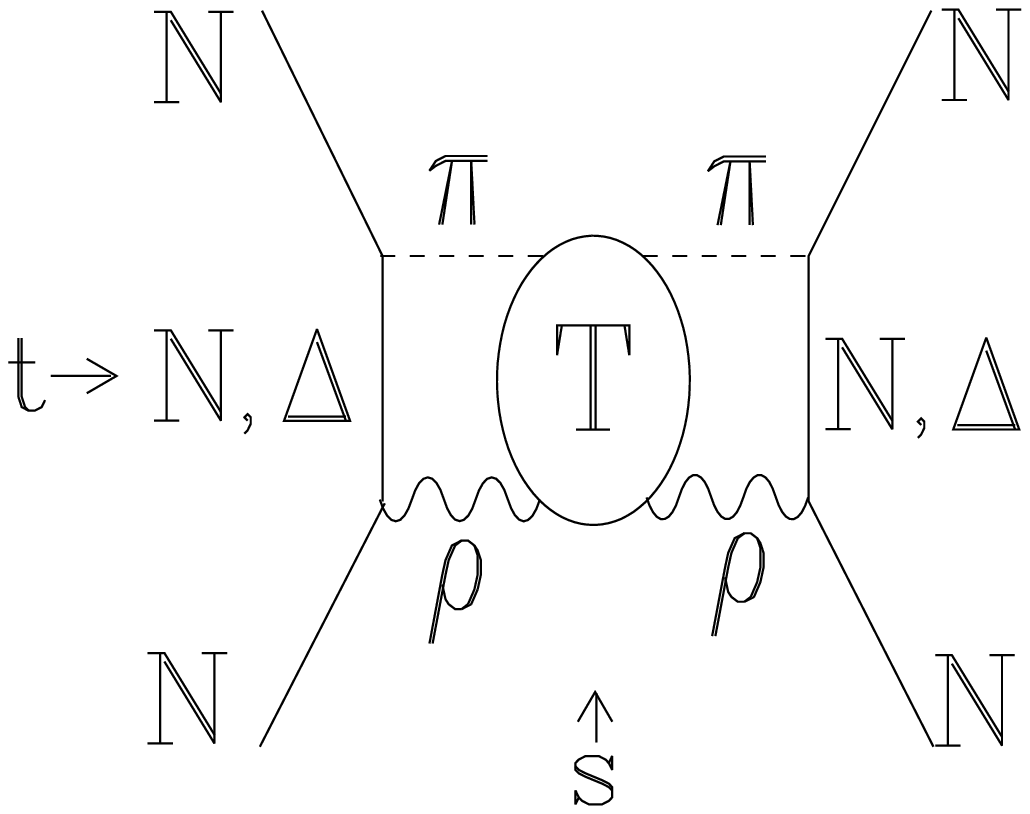}
\hbox to \hsize{\hskip2.8truecm{\noindent
\sl Fig.9: }\vtop{\advance\hsize by -5.6truecm\advance\hsize by -30pt
\noindent \sl Correlated $\pi\rho$ exchange.}\hfill}
\vskip0.5cm

\noindent
Again the result can be represented as an integral over various spectral
functions in the different channels $J^P(I^G) = 0^-(1^-)$ ("$\pi$"),
$1^-(0^-)$ ("$\omega$"), $1^+(1^-)$ ("$A_1$"), $1^+(0^-)$ ("$H_1$")
which we have considered. Thus we
have for the pionic channel:
$$
V^{(0^-)}_{N,N;N,N} (t) \sim
\int\limits^{\infty}_{(m_{\pi}+m_{\rho})^2} dt' \,
{ {\rho^{_(0^-)}(t')} \over {t'-t} }  \; .
\eqno(4)
$$
$\rho^{(0^-)}$ is shown in Fig.10. Obviously it provides a sizable
contribution,
with a peak around 1.1~GeV, somewhat smaller than the mass (1.2 GeV) of
the phenomenological $\pi'$ introduced in our recent $NN$ models to
accommodate a soft $\pi NN$ form factor [12].


\midinsert
\phantom{ }
\vskip 5.25cm
\includegraphics{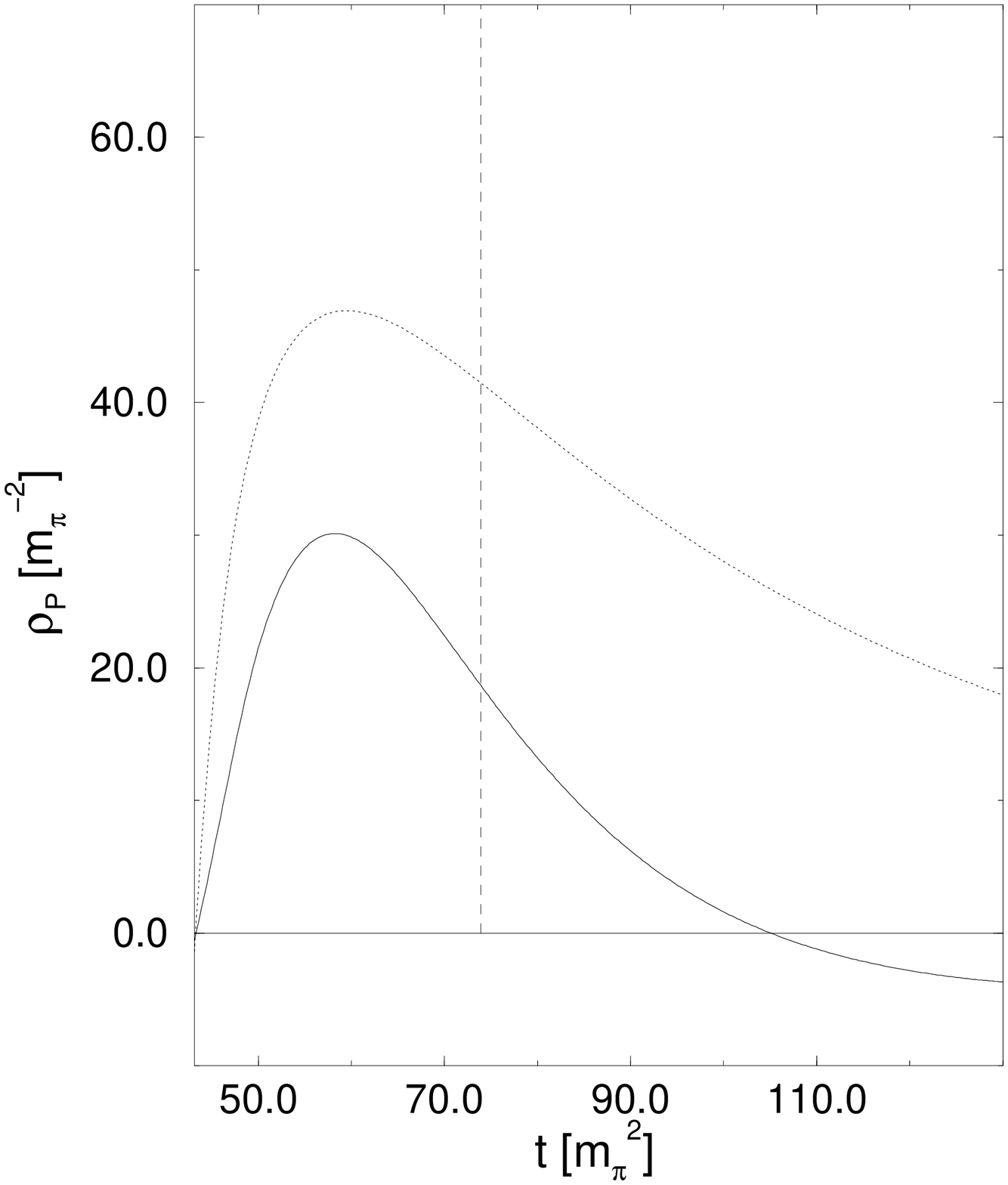}
\hbox to \hsize{\hskip2.8truecm{\noindent
\sl Fig.10: }\vtop{\advance\hsize by -5.6truecm\advance\hsize by -35pt
\noindent \sl The spectral function in the pionic channel
$\rho^\pi_P(t)$. The dotted line shows the uncorrelated part whereas the
solid line represents the correlated contribution. The vertical dashed
line represents sharp mass $\pi'$ exchange used in
Ref.[12].}\hfill}
\endinsert

\noindent
Indeed as shown in Fig.11, the resulting interaction due to correlated
$\pi \rho$ exchange in the pionic channel is able to counterbalance the
substantial suppression induced in the one pion-exchange potential
(OPEP) when going from a cutoff mass
$\Lambda_{NN \pi}$ of 1.3 GeV, phenomenologically required in the
(full) Bonn
potential, to a value of 1.0 GeV. For basic theoretical reasons such a
reduction is highly welcome, since various information from other
sources point to a rather soft $\pi NN$ form factor characterized by
$\Lambda_{NN \pi} \cong 0.8$ GeV [13].
This value will probably be reached if correlated $\pi \sigma$
exchange is included, too, which is also missing in the Bonn potential.
(As usual, $\sigma$ stands for a low mass correlated $\pi \pi$ pair in
the $0^+$
channel). Thus it appears that in the Bonn potential OPEP
together with a hard form factor is an effective
description of 'true' one-pion exchange (with a soft form factor) plus
correlated $\pi \rho$ (and $\pi \sigma$) exchange in the pionic
channel.

\midinsert
\phantom{ }
\vskip 4.75cm
\includegraphics{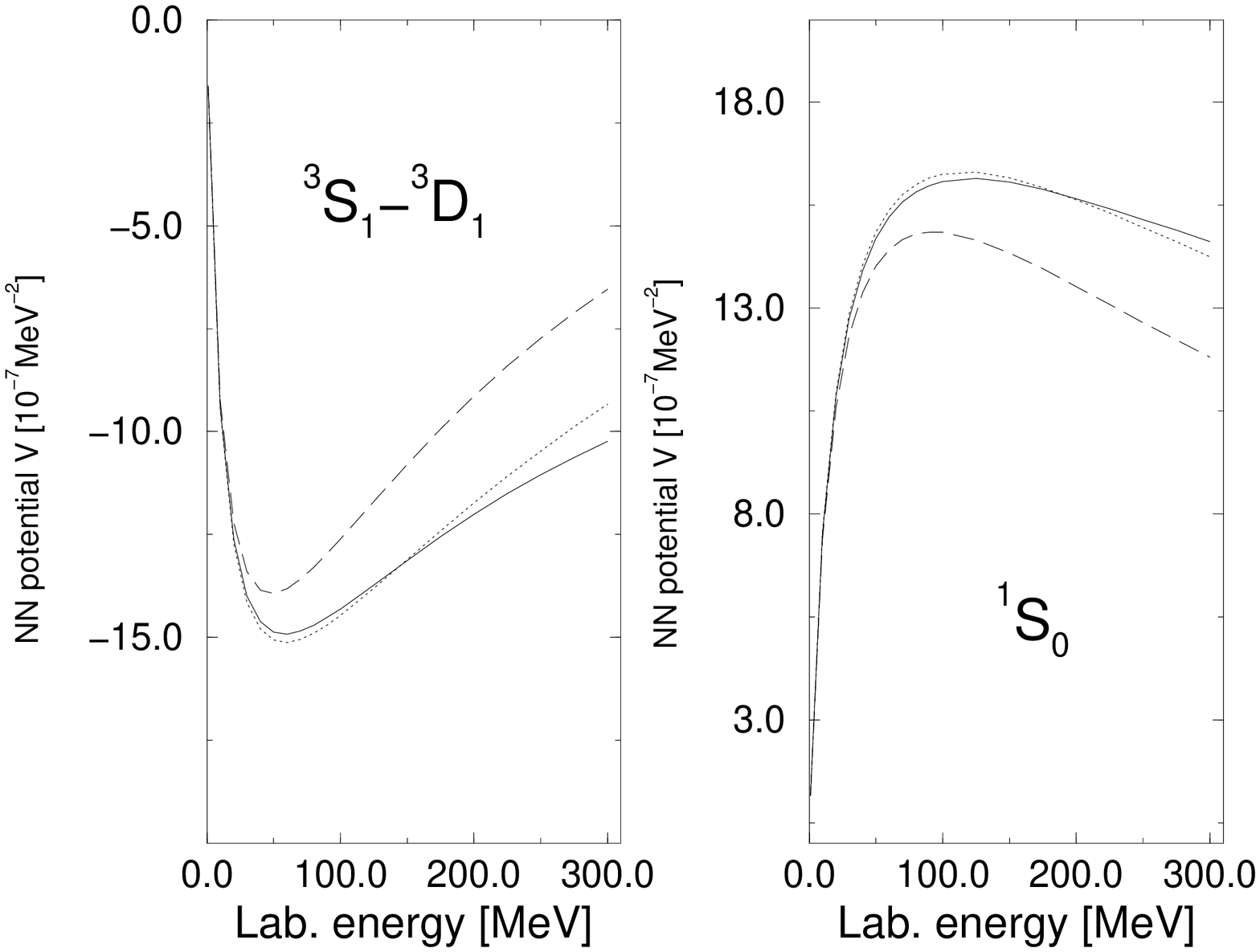}
\hbox to \hsize{\hskip2.8truecm{\noindent
\sl Fig.11: }\vtop{\advance\hsize by -5.6truecm\advance\hsize by -35pt
\noindent
\sl On-shell $NN$ potential $V_{NN}$ as function of the
nucleon lab energy for the $^3D_1\,-\,^3S_1$ transition (lefthand
side) and in the $^1S_0$ state (righthand side).  The dotted line
denotes the one-pion-exchange potential as used in the Bonn potential
($g^2_{\pi NN}/4\pi=14.4,\,\Lambda_{\pi NN}= 1.3\,GeV$). For the
dashed line, $\Lambda_{\pi NN}=1.0\,GeV$ was used. The solid line
results if correlated $\pi\rho$ exchange in the pionic channel is
added to the dashed line.}\hfill}

\endinsert

\noindent
In the $\omega$ channel, the exchange of a correlated $\pi \rho$ pair
also provides a sizable contribution to the $NN$ interaction. Here we
have included the genuine $\omega$ meson explicitly and replaced the
(effective) $\omega$ exchange in the Bonn potential by the resulting
correlated $\pi \rho$ potential, which can be decomposed into a pole
and a non-pole piece.  The former provides a microscopic model for
'true' $\omega$ exchange leading to a renormalized $\omega NN$
coupling constant at the pole, $g^2_{NN \omega}/4 \pi$ = 11.0, which
is about a factor of 2 smaller than the effective value of 20 used in
the Bonn potential. Thus 'true' correlated $\pi \rho$ exchange, i.e.\
the non-pole piece, provides almost half of the empirical repulsion
needed in the $NN$ interaction. It can be roughly parametrized by
sharp-mass $\omega'$-exchange with $g^2_{NN \omega}/4\pi$ = 8.5,
$f_{NN \omega}/g_{NN \omega}$ = 0.4 and $m_{\omega'}$ =1120 MeV.

\noindent
The new reduced coupling constant (11.0) is still about a factor of 2
larger than provided by customary SU(3) estimates, which use
$g^2_{NN \omega} = 9 g^2_{NN \rho}$. Thus with
$g^2_{NN \rho}/4\pi$ = 0.55 as determined by Hoehler and
Pietarinen [14] we have $g^2_{NN \omega}/4\pi$ = 5. Note however,
that the above relation
between $\omega$ and $\rho$ coupling constants is based, apart from
ideal mixing,
on the assumption of vanishing $\phi NN$ coupling. For $g_{NN \phi}$
unequal to zero the above relation goes into
$$
g_{NN \omega} = 3 g_{NN \rho} - \sqrt{2} g_{NN \phi} \; .
\eqno(5)
$$
If we take $g_{NN \phi} = - g_{NN \rho}$, (which amounts to a rather
small deviation from
zero) we have $g^2_{NN \omega} \approx 20 g^2_{NN \rho}$, in rough
agreement with our results.
Such a value for the $\phi NN$ coupling to the nucleon and the negative
sign is quite conceivable from the $\phi$ coupling to the nucleon via
the $\Lambda \bar \Lambda$ continuum [15].
\bigskip
\medskip
\goodbreak
\noindent
\leftline{\hskip 2.8 truecm  {\bf MESON-MESON CORRELATIONS IN
} }
\smallskip
\leftline{\hskip 2.8 truecm  {\bf NUCLEON-ANTINUCLEON ANNIHILATION
} }\medskip
\noindent
We now want to turn our attention to the role of meson-meson
correlations in nucleon-antinucleon annihilation. Reactions involving
antinucleons have always been considered to be the ideal place for
finding quark effects since annihilation phenomena from the
$N \bar N$ system
are supposedly governed by short-distance physics. Therefore the basic
question is: Do  these phenomena proceed  via baryon exchange
(Fig.12a) or via quark-gluon exchange (Fig.12b,c)?

\phantom{ }
\vskip 3.7cm
\includegraphics{fig12.eps}
\hbox to \hsize{\hskip2.8truecm {\noindent
\sl Fig.12: }\vtop{\advance\hsize by -5.6truecm\advance\hsize by -35pt
\noindent \sl Baryon exchange (a) and quark-gluon (b,c) transition mechanism
for the annihilation process $N\bar N\to M_1 M_2$.}\hfill}
\vskip0.5cm

\noindent
As usual this question must
be decided not by our personal preferences but by a serious
confrontation
with the empirical data. Unfortunately the data do not only see the
bare
transition mechanisms but also the sum of various higher order
processes
involving initial and final state interactions (see Fig.13). In
present-day studies initial state interaction effects are usually
taken into
account whereas effects from final-state meson-meson correlations are
in general neglected simply because not much is known about
meson-meson
interactions. Based on our studies of the foregoing chapters we have
now the possibility to investigate the role of $\pi \pi$ (and
$K \bar K$) as well
as
$\pi\rho$ correlations in $p \bar p$ annihilation.

\phantom{ }
\vskip 3.80cm
\includegraphics{fig13.eps}
\hbox to \hsize{\hskip2.8truecm{\noindent
\sl Fig.13: }\vtop{\advance\hsize by -5.6truecm\advance\hsize by -35pt
\noindent \sl The transition amplitude with its contributions from the
Born diagram (a), initial (b,d) and final state interactions (c,d).}\hfill}
\bigskip
\goodbreak
\medskip
\noindent
{\bf \b{p \bar p \rightarrow  \pi \pi, K \bar K, \rho \rho} }
\smallskip\noindent
Our calculations are based on a coupled channel model involving the
channels $p \bar p$, $\pi \pi$, $K \bar K$, and $\rho \rho$. After
specifying all diagonal
and transition interactions one obtains in this way a simultaneous
description of the elastic ($p \bar p \rightarrow p \bar p$) and
annihilation $p \bar p \rightarrow \pi \pi$,
$p \bar p \rightarrow K \bar K$, $p \bar p \rightarrow \rho \rho$
reactions.

\phantom{ }
\vskip 7.5cm
\includegraphics{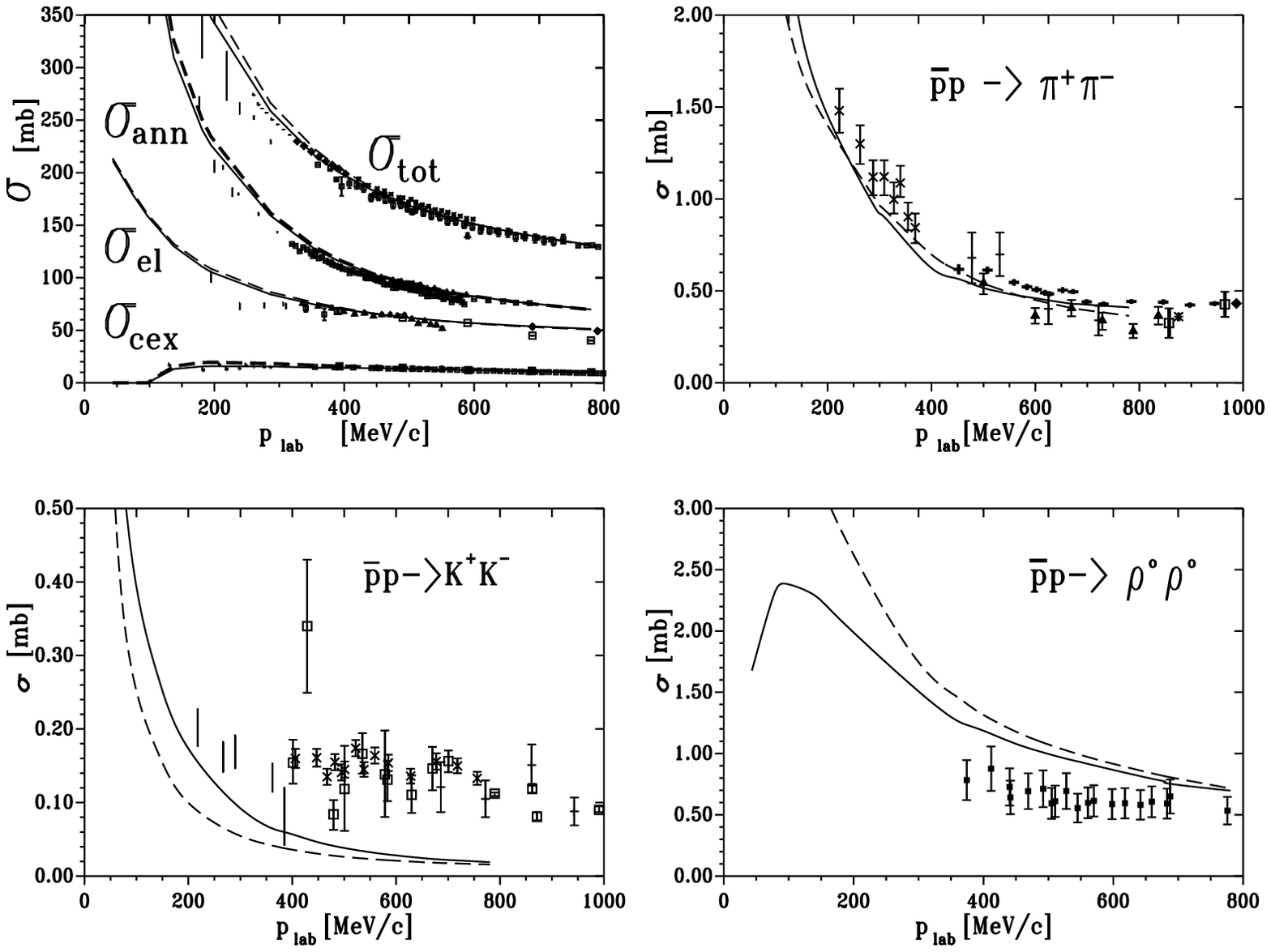}
\hbox to \hsize{\noindent\hskip2.8truecm
{\noindent\sl Fig.14: }\vtop{\advance\hsize by -5.6truecm\advance\hsize by
-35pt
\noindent \sl Cross sections for $N \bar N$ scattering as well as
annihilation from the $N\bar N,\pi\pi,K\bar K,\rho\rho$ coupled
channels system. The results of the full calculation are denoted by
the solid curves. For the dashed curves, meson-meson correlations have
been omitted. For references to the empirical data, see
Ref.[16].}\hfill}
\vskip0.5cm

\noindent
Fig.14 shows the resulting cross sections. Obviously the model gives
a reasonable account of the empirical situation. Furthermore, meson-
meson correlations play only a very small role. However, when we go
to differential cross sections and the analyzing power for
$p \bar p \rightarrow \pi \pi$
where empirical data are available, we notice sizable effects due to
final state (meson-meson) correlations, see Fig.15.

\phantom{ }
\vskip 3.45cm
\includegraphics{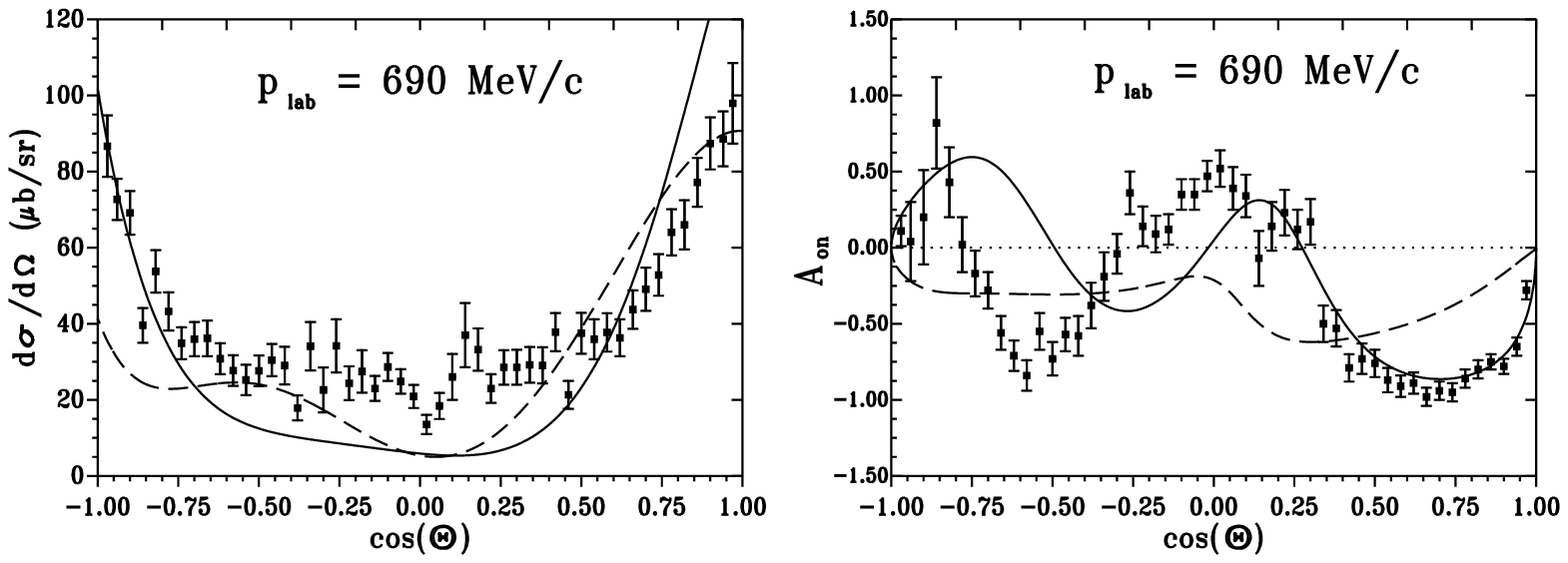}
{

\hbox to \hsize{\hskip 2.8truecm
{\noindent\sl Fig.15: }\vtop{\advance\hsize by -5.6truecm \advance\hsize by
-35pt
\noindent \sl Differential cross section and analyzing
power
for the $N \bar N\to \pi^+\pi^-$ annihilation process
at $p_{lab}=690MeV/c$.
The curves have the same meaning as in Fig.14. For references of
the empirical data, see Ref.[16].}\hfill}
}
\goodbreak
\medskip
\noindent
{\bf \b{p \bar p \rightarrow \pi \rho} }
\smallskip\noindent
Here [17] we have done a Distorted-Wave-Born-Approximation (DWBA)
calculation
with the initial state interaction (ISI)
$p \bar p \rightarrow p \bar p$ taken from Ref. [16] and
the final $\pi \rho \rightarrow \pi\rho$ interaction (FSI) from
Ref. [9]. We obtain for partial
cross sections from specific $p \bar p$ initial S-states results
shown in
Table~2. The inclusion of the initial state cuts down the Born results
by an order of magnitude while the inclusion of final state correlation
effects has a comparably minor effect only. However the direction of
effects is different  in the two states: For $^3S_1$ the cross section
is
increased while for $^1S_0$ it is decreased. This different behaviour
is
just due to a different overall sign of the $\pi\rho$ amplitude
in the different isospin states. Consequently final state
interaction effects induce a drastic increase in the ratio,
in agreement with the large empirical value, and thus are able
to solve the '$\pi\rho$ puzzle'.
%
%
\midinsert
\hbox to \hsize{\noindent\hskip 2.8truecm
{\sl Table 2: }\vtop{\advance\hsize by -5.6truecm \advance\hsize by -35pt
\noindent \sl Partial cross sections for $\bar p p \rightarrow \pi \rho$
in [$\mu$b] at $p_{lab}$ = 100 MeV/c.}\hfill}
$$
\vbox{
{\offinterlineskip \tabskip=0pt
\halign{  \strut
	#\quad&
          #  &
          #&
          \quad\hfill #\quad &
          #&
          \quad\hfill #\quad &
          #&
          \quad \hfill # \quad &
          #&
          \quad \hfill # \quad &
#
          \cr
\noalign{\hrule}
&    && Born && + ISI  && + FSI && experiment & \cr
\noalign{\hrule}
& $\bar p p (^3S_1,I=0)$ && 435 && 3.40 && 4.42 &&             & \cr
& $\bar p p (^1S_0,I=1)$ && 126 && 0.73 && 0.14 &&             & \cr
\noalign{\hrule}
& Ratio R                && 3.5 && 4.7  && 32   && 35 $\pm$ 16 & \cr
\noalign{\hrule}
}}}$$
\endinsert
%
%
\noindent
\leftline{\hskip 2.8 truecm  {\bf SUMMARY} }
\smallskip
\nobreak
\noindent
As a main conclusion, meson-meson correlations are important in
baryon-baryon as well as anti\-baryon-baryon interactions and have to
be included in any realistic model description.

\noindent
In the baryon-baryon ($BB$) sector, an explicit dynamical model for
correlated $\pi \pi$ (plus $K \bar K$) exchange leads to results
characteristically
different from the simplified treatment in terms of sharp-mass
$\sigma'$
and $\rho$ exchange; for example, the effective strengths do not fulfil
SU(3) relations. In the scalar-isoscalar channel the predicted
attraction decreases with increasing strangeness (becoming more
negative) of the $BB$ system.

\noindent
Furthermore a model for correlated $\pi\rho$ exchange between two
nucleons
has been discussed, which has substantial effects: In the pionic
channel
it counterbalances the suppression generated by a soft $\pi NN$ form
factor
of monopole type with a cutoff mass of about 1 GeV; in the
$\omega$-channel
it provides a strong repulsive contribution, leaving little room for
explicit quark-gluon effects.

\noindent
The same meson-meson correlations have been discussed in the $N \bar N$
sector.
They have substantial effects in $p \bar p \rightarrow \pi \pi$ spin
observables and
$p \bar p \rightarrow \pi\rho$ partial cross sections, in both cases
improving the
agreement
with experiment. Therefore such correlations have to be included before
the relevant transition mechanisms (baryon exchange on one hand and
quark rearrangement or quark annihilation on the other hand) can be
reliably identified.

\vfil

\bigskip
\leftline{\hskip 2.8 truecm  {\bf REFERENCES} } \medskip
%
%
%

\hang\noindent{\phantom {1}[1] R. Machleidt, K. Holinde, and
Ch. Elster, (1987). {\sl Phys. Rep.}, {\bf 149}, 1.}

\hang\noindent{\phantom {1}[2] B. Holzenkamp, K. Holinde, and J. Speth,
(1989). {\sl Nucl. Phys.}, {\bf A500}, 485.}

\hang\noindent{\phantom {1}\phantom{[2]} A. Reuber, K. Holinde, and
J. Speth, (1994). {\sl Nucl. Phys.}, {\bf A570}, 543.}

\hang\noindent{\phantom {1}[3] K.F. Liu, S.J. Dong, and W. Wilcox,
(1995). {\sl Phys. Rev. Lett.}, {\bf 74}, 2172.}

\hang\noindent{\phantom {1}[4] J. Haidenbauer and K. Holinde, (1989).
{\sl Phys. Rev.}, {\bf C40}, 2465.}

\hang\noindent{\phantom {1}[5] H.-C. Kim, J. W. Durso, and K. Holinde,
(1994). {\sl Phys. Rev.}, {\bf C49}, 2355.}

\hang\noindent{\phantom {1}[6] A. Reuber, (1995).
{Berichte des Forschungszentrums J\"ulich, Nr. 3117.}}

\hang\noindent{\phantom {1}\phantom{[6]} \negthinspace
A. Reuber, K. Holinde, H.-C. Kim, and
J. Speth, Correlated $\pi \pi$ and $K \bar K$ Exchange in the
Baryon-Baryon Interaction, submitted to {\sl Nucl. Phys.}, {\bf A}.}

\hang\noindent{\phantom {1}[7] G. H\"ohler, F. Kaiser, R. Koch, and
E. Pietarinen, (1979).
"Handbook of Pion-Nucleon Scattering",
{\sl Phys. Data 12-1}, Fachinformationszentrum, Karlsruhe}

\hang\noindent{\phantom {1}\phantom{[7]} \thinspace O. Dumbrajs et al.,
(1983).
"Compilation of Coupling Constants and Low-Energy Parameters",
{\sl Nucl. Phys.}, {\bf B216}, 277.}

\hang\noindent{\phantom {1}[8] R.A.M. Klomp, V.G.J. Stoks, and
J.J. de Swart, (1991). {\sl Phys. Rev.}, {\bf C44}, R1258.}

\hang\noindent{\phantom {1}[9] G. Janssen, K. Holinde, and J. Speth,
(1994). {\sl Phys. Rev.}, {\bf C49}, 2763.}

\hang\noindent{[10] D. Lohse, J.W. Durso, K. Holinde, and J. Speth,
(1990). {\sl Nucl. Phys.}, {\bf A516}, 513.}

\hang\noindent{[11] C. Schuetz, K. Holinde, J. Speth, B.C. Pearce,
and J.W. Durso, (1990). {\sl Phys. Rev.}, {\bf C51}, 1374.}

\hang\noindent{[12] J. Haidenbauer, K. Holinde, and A. W. Thomas,
(1994). {\sl Phys. Rev.}, {\bf C49}, 2331.}

\hang\noindent{[13] S.A. Coon, and M.D. Scadron, (1990).
{\sl Phys. Rev.}, {\bf D42}, 2256.}

\hang\noindent{[14] G. H\"ohler and E. Pietarinen, (1975).
{\sl Nucl. Phys.}, {\bf B95}, 210.}

\hang\noindent{[15] V. Mull, privat communication.}

\hang\noindent{[16] V. Mull and K. Holinde, (1995).
{\sl Phys. Rev.}, {\bf C51}, 2360.}

\hang\noindent{[17] V. Mull, G. Janssen, J. Speth, and K. Holinde,
(1995). {\sl Phys. Lett.}, {\bf B347}, 193.}
\bye